\documentclass[useAMS,usenatbib]{mn2e} \usepackage{graphicx}
\usepackage{longtable,lscape} \usepackage{amssymb}
\usepackage[usenames]{color}

\def\aap{A\&A}  \def\apss{Ap\&SS} \def\apj{ApJ}
\def\apjs{ApJS} \def\mnras{MNRAS}

\title[Study of sdO models: mode trapping]{Study of sdO models: mode
  trapping} \author[C. Rodr\'\i guez-L\'opez, A. Moya, R. Garrido,
  J. MacDonald, R. Oreiro and A. Ulla]{C. Rodr\'\i
  guez-L\'opez$^{1,2,3}$\thanks{E-mail:crodrigu@ast.obs-mip.fr}
  A. Moya$^{3,4}$, R. Garrido$^{3}$, J. MacDonald$^{5}$,
  R. Oreiro$^{6}$ and A. Ulla$^{2}$ \\
  $^{1}$Laboratoire
  d'Astrophysique de Toulouse-Tarbes, Universit\'e de Toulouse, CNRS,
  Toulouse 31400, France \\ $^{2}$Departamento de F\'\i sica Aplicada,
  Universidade de Vigo, Vigo 36310, Spain \\ $^{3}$Departamento de
  F\'isica Estelar, Instituto de Astrof\'\i sica de Andaluc\'\i
  a-CSIC, Granada 18008, Spain\\ $^{4}$Laboratorio de Astrof\'isica
  Estelar y Exoplanetas, LAEX-CAB (INTA-CSIC), PO BOX 78, 28691
  Villanueva de la Ca\~nada, Madrid, Spain \\ $^{5}$University of
  Delaware, Department of Physics and Astronomy, Newark, DE 19716, USA
  \\ $^{6}$Institute of Astronomy, Katholieke Universiteit Leuven,
  Celestijnenlaan 200D, 3001 Leuven, Belgium}

\begin{document}

\date{Accepted 2009 Month dd. Received 2009 Month dd; in original form
  2009 Month dd}

\pagerange{\pageref{firstpage}--\pageref{lastpage}} \pubyear{2009}

\maketitle

\label{firstpage}

\begin{abstract}

We present the first description of mode trapping for sdO models. Mode
trapping of gravity modes caused by the He/H chemical transition is
found for a particular model, providing a selection effect for high
radial order trapped modes. Low- and intermediate-radial order {\em
  p}-modes (mixed modes with a majority of nodes in the P-mode region)
are found to be trapped by the C-O/He transition, but with no
significant effects on the driving. This region seems to have also a
subtle effect on the trapping of low radial order {\em g}-modes (mixed
modes with a majority of nodes in the G-mode region), but again with
no effect on the driving. We found that for mode trapping to have an
influence on the driving of sdO modes (1) the mode should be trapped
in a way that the amplitude of the eigenfunctions is lower in a
damping region and (2) in this damping region significant energy
interchange has to be produced.
\end{abstract}


\section{Introduction}
Subdwarf O (sdO) stars cover a domain of approximately 60\,000~K in
$T_\mathrm{eff}$ and 2.5~dex in $\log g$ in the hot and high gravity
domain of the HR diagram. Due to the wide spread of physical
parameters their evolutionary stage is not completely understood, but
some progress has recently been made with larger than ever (comprising
up to a hundred sdOs) spectroscopic studies (\citealt{stroer07};
\citealt{hirsch08}). Determinations of $T_\mathrm{eff}$, $\log g$ and
helium abundance allowed \citet{stroer07} to make some statistical
correlations between position in the HR and evolutionary stage, and as
a result they propose that sdOs come from a mixed population of
post-EHB (i.e. sdB descendants), post-AGB and post-RGB objects. The
canonical picture depicts them as objects of about half a solar mass
with an inert carbon-oxygen core, and helium and hydrogen burning
layers.

\begin{table*}
\centering
\caption{Main physical parameters and mass fractions of the sdO model
  used. X(other) refers to the mass fractions of all the other
  elements. Z is the initial and current metallicity of the model.}
\label{tab:model}
\setlength{\tabcolsep}{4pt}
\begin{tabular}{l|cccccccccccc}  
\hline $T_\mathrm{eff}$ & $\log g$ & M & $\eta_\mathrm{R}$ & X(H) &
X(He$^3$) & X(He$^4$) & X(C) & X(N) & X(O) & X(other) & Z \\ (K) & &
(M$_\odot$) & & & & & & & & & \\ \hline 45\,000 & 5.26 & 0.469 & 0.650
& 0.59 & 1.2E-03 & 0.35 & 7.4E-03 & 4.2E-03 & 2.4E-02 &1.5E-02 & 0.05
\\ \hline
\end{tabular}
\centering
\end{table*}

A milestone for sdOs was passed in 2006 with the discovery of the
first, and unique to date, pulsating sdO SDSS\,J160043.6+074802.9
(known in short as J1600+0748) by \citet{woudt06}. This object is a
very fast pulsator, with a main period of $\sim$120~s, and at least
seven other frequencies \citep{crl09sdss} between the main frequency
and its first harmonic. The excited frequencies were identified by
\citet{fontaine08} as {\em p}-modes and its pulsations explained as a
classical $\kappa$ mechanism with the aid of appropriate models
including radiative levitation of iron. Our own attempts of exciting
{\em p}-mode pulsations in uniform metallicity sdO models failed
(\citealt{crl09}, Paper~I), although some models, as this one, were
found to present a certain tendency to instability in the {\em p}-mode
region. During the course of our investigation of pulsational
properties of sdO models, for which {\em g}-mode driving was found, a
particular model presenting {\em g}-mode trapping was discovered.

In main sequence stars the asteroseismic potential of {\em g}-mode
trapping to study the mixing processes within the convective core and
their effects on the surrounding chemical gradient have been well
documented by \citet{miglio08}. In the case of compact pulsators, {\em
  g}-mode trapping has been described extensively for sdBs
\citep{charpinet00}, DA and DB white dwarfs, and GW Vir pre-white
dwarfs (see e.g. \citealt*{winget81}; \citealt{brassard91};
\citealt{brassard92}, or more recently \citealt{corsico06};
\citealt{gautschy02}) as due to chemical stratification in the
envelope of these stars. Because gravity modes are trapped in the
outer envelope, they have lower amplitude in the damping core, which
intensifies their instability. This may help, in some cases, alleviate
the problem of observing less modes than predicted by the models. It
also provides a useful tool to determine the mass and width of the
surface layer (see e.g. \citealt{brassard92};
\citealt{kawaler94}). \citet{charpinet00} reported gravity mode
trapping due to the transition between the helium core and the
hydrogen-rich envelope in sdB models. They also identified subtle, but
non-negligible departures from uniform frequency spacing for pressure
modes, caused by the same chemical transition region.

In sdB models, the maximum gradient in Brunt-V\"ais\"al\"a (BV)
frequency is produced at the transition of the helium radiative core
to the hydrogen envelope. In sdO models the BV profile is more
complex, since as evolution proceeds, a C-O core builds up, resulting
in two chemical transitions: one from the C-O core to the He burning
shell, and another one from the He radiative shell to a H burning
shell. This renders the mode trapping effects more complex in sdO than
in sdB models.

We report in this study the discovery of {\em g}-mode trapping in a
sdO model caused by the He/H transition. This provides a selection
mechanism in the way that trapped modes would be more easily
excited. Low- and intermediate radial order {\em p}-modes\footnote{We
  remind the reader that in sdO models, modes with low to intermediate
  frequencies behave as mixed modes, i.e. have nodes both in the P-
  and G-mode region. The classification 'pressure modes' refers then
  to mixed modes with most of their nodes in the P-mode region (see
  fig.~7 of Paper~I).} are found to be trapped by the deeper
transition from the C-O/He, but with no significant effects on the
driving.

The importance of {\em g}-mode trapping is evident as a way to probe
the deep stellar interior, where gravity modes propagate, unattainable
in any other way; but also to derive information about the location
and width of the chemical transitions and the mass of the envelope.

The paper is organized as follows: in Section~2 we describe some
general characteristics of the sdO model used. In Section~3 we
describe the {\em g}-mode trapping and the effects of cancelling out
the He/H and C-O/He chemical transitions in BV frequency on the
trapping and driving of the modes. Section~4 does a similar treatment
for {\em p}-modes. Finally, Section~5 presents a summary and Section~6 the
discussion and conclusions.

\section{The Model}
\begin{figure}
   \includegraphics[width=9cm]{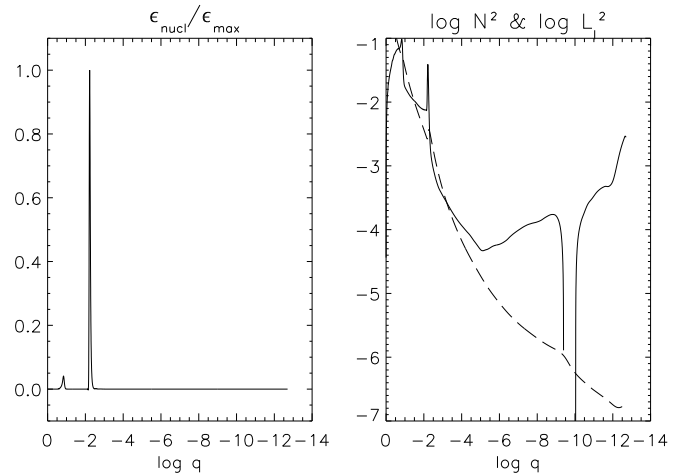}
   \caption{Left: Normalised nuclear energy generation rate as a
     function of the fractional mass depth parameter ($\log q = 1- M_r
     / M_T$). The deeper peak at $\log q \simeq -0.8$ corresponds to
     the helium burning shell; the shallower peak at $\log q \simeq
     -2.2$ to the hydrogen burning shell. Right: Brunt-V\"ais\"al\"a
     frequency (solid line) showing the two peaks corresponding to the
     maximum chemical gradients: from the C-O core to the He burning
     shell ($\log q \simeq -0.8$), and from the He radiative shell to
     hydrogen shell burning ($\log q \simeq -2.2$). Lamb frequency
     (dashed line) for $\ell=2$ modes is also shown.}
   \label{fig:phys2epsn2}
   \end{figure}

This theoretical exercise arose as an exploration of a particular sdO
model, whose properties were thoroughly described in
Paper~I\footnote{The model corresponds to model 8, also named
  eta\,650t45, according to its mass loss rate parameter and effective
  temperature.} and parameters given in Table~\ref{tab:model}. The
model, built with {\scriptsize JMSTAR} code \citep{lawlor06}, comes
from a 1~M$\odot$ star on the pre-main sequence with enhanced mass
loss rate on the red giant branch, that drives the star to evolve to
higher temperatures at constant luminosity. A delayed helium-flash
puts the star back in the horizontal branch, and further evolution
brings the star to the sdO domain. The sdO models have developed a
carbon-oxygen core while helium and hydrogen shell burning is still
produced (Fig.~\ref{fig:phys2epsn2} left). The change in chemical
composition from the C-O core to the He burning shell, and from the He
radiative layer to the H burning shell is seen in the two steep peaks
in BV frequency and subtler transitions in Lamb frequency
(Fig.~\ref{fig:phys2epsn2} right).

The model, which was subject to a non adiabatic analysis with
{\scriptsize GRACO} (\citealt{moya04}; \citealt{moya08}), was found
stable in the frequency range $\sim$0.5 to 25~mHz. However, an
oscillatory behaviour of the growth rate at low and intermediate
frequencies (see fig.~10 of Paper~I) caught our attention. That drove
us to explore the behaviour of the kinetic energy of the modes, which
revealed that the model experienced mode trapping effects.

\section{Gravity modes}

 \begin{figure}
   \includegraphics[width=9cm]{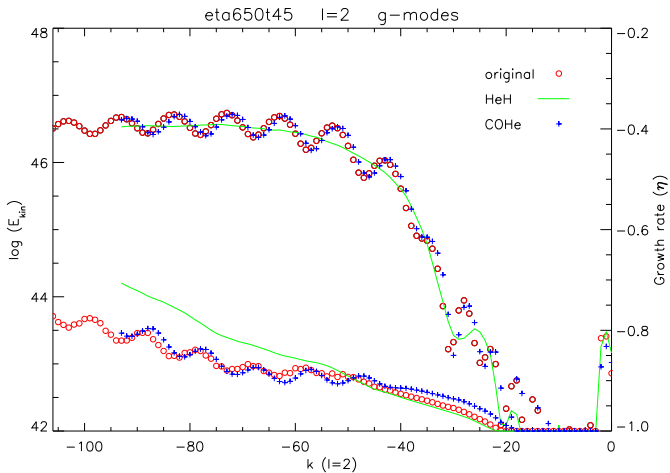}
   \caption{Logarithm of the total kinetic energy vs. radial order for
     {\em g}-modes (this and subsequent plots have been made for
     $\ell=2$ modes) of the original (upper circles) and perturbed models
     without the He/H (upper solid line) and C-O/He (upper crosses) transition
     regions. Lower symbols plot the corresponding growth rates whose
     values can be read on the right axis.}
   \label{fig:allgkin}
   \end{figure}

   \begin{figure}
   \includegraphics[width=9cm]{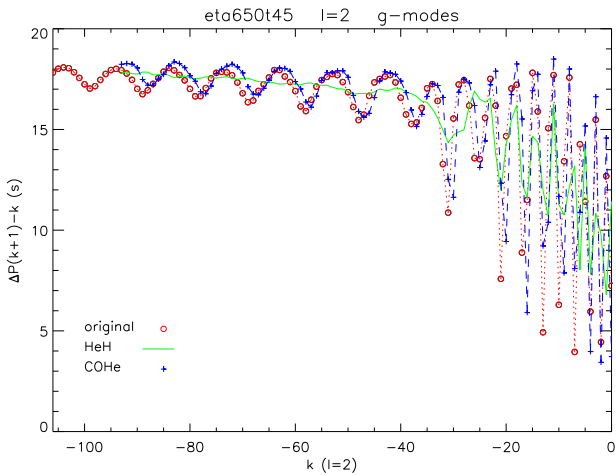}
   \caption{Period spacing for {\em g}-modes with the same degree
     $\ell$ and consecutive radial orders for the original (circles
     and dotted line) and perturbed He/H (solid line) and C-O/He
     (crosses and dashed line) models.}
   \label{fig:allgdeltap}
   \end{figure}

The logarithm of the total kinetic energy of each {\em g}-mode and its
growth rate $\eta$ are given in Fig.~\ref{fig:allgkin} (upper and
lower circles, respectively, the meaning of the other symbols will be
explained below). The plot reveals that in the region of very
high-radial order {\em g}-modes (radial order $k \gtrsim 40$), there
are small variations in the kinetic energy, of no more than one order
of magnitude, between different modes. We also see that modes with
local minimum values of the kinetic energy have local maximum growth
rates and vice versa.

The non-uniform distribution of the kinetic energy resembles the mode
trapping phenomenon caused by the potential barrier due to the
composition transition regions. The chemical transition regions act as
resonant cavities for the modes, changing the amplitudes of the
eigenfunctions in different regions of the star and hence modifying
their kinetic energy, which is given by:

\begin{equation}
E_{kin} = \frac{1}{2} \sigma^2 \int_0 ^R \left( |\xi_r|^2 +
l\left(l+1\right) |\xi_h|^2 \right) 4 \pi \rho r^2 dr
\label{eq:ekin-mode}
\end{equation}

\noindent where $\sigma$ is the eigenfrequency, $\xi_r$ and $\xi_h$
the radial and horizontal displacement eigenfunctions, and $\rho$ the
local density. It follows that {\em g}-modes will have higher kinetic
energies than {\em p}-modes, as they propagate in denser regions of
the star. This is clearly seen by comparing Fig.~\ref{fig:allgkin} and
the corresponding plot for p-modes, Fig.~\ref{fig:allpkin}.

Following the asymptotic theory of nonradial oscillations (see
e.g. \citealp{tassoul80}; \citealp{smeyers95}; \citealp{smeyers07}),
modes of the same degree $\ell$ and consecutive radial order $k$ would
have a uniform period separation, $\Delta P$, given by:

\begin{equation}
\Delta P = P_{k+1,l} - P_{k,l}= 2 \pi^2 \left[ \int_0 ^R \frac{|N|}{r}
  dr \right]^{-1} \left[l \left(l+1\right)\right]^{- \frac{1}{2}}
\label{eq:separeixon}               
\end{equation}

\noindent where $N$ is the BV frequency. Thus, mode trapping effects
are revealed for certain modes as deviations with respect to the mean
period spacing. This can be seen from the data points plotted in red
in Fig.~\ref{fig:allgdeltap}.

\begin{figure*}
     \begin{tabular}{cc}
 \resizebox{0.45\linewidth}{6cm}{\includegraphics[angle=90]{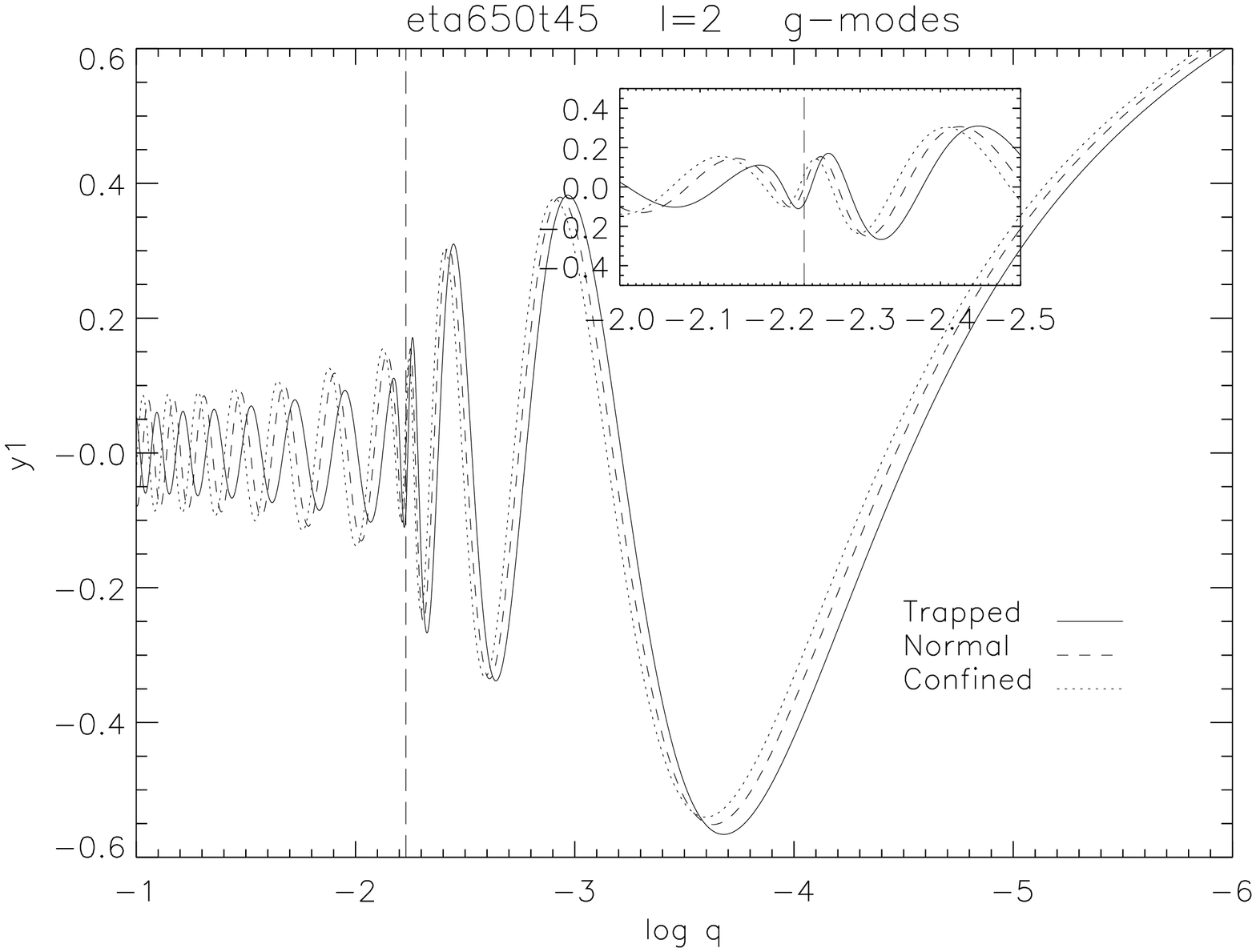}}
 &
 \resizebox{0.45\linewidth}{6cm}{\includegraphics[angle=90]{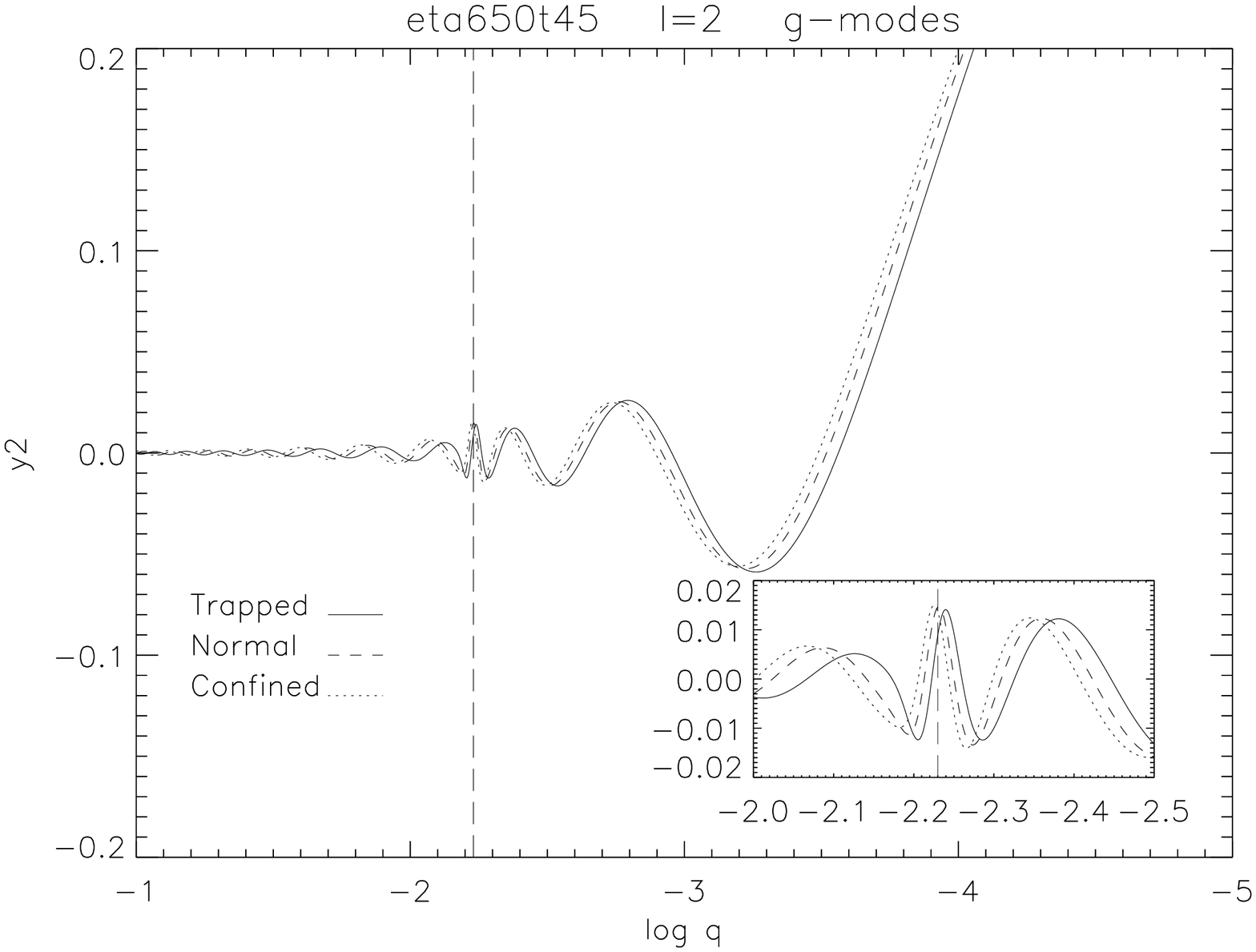}}
 \\
      \end{tabular}
   \caption{Radial (left) and horizontal (right) displacement
     eigenfunctions for trapped ({\em g}88), normal ({\em g}85) and
     confined ({\em g}83) {\em g}-modes of the original model. The
     vertical line marks the location of the He/H transition. The
     insets show a blow-up of this region.}
   \label{fig:gradial88}
   \end{figure*}

The radial and horizontal displacement eigenfunctions (y$_1=\xi_r/r$
and y$_2=(\sigma^2 r/g) * \xi_h/r$, respectively as defined by
\citealt{dziembowski71}) of representative modes with minimum ({\em
  g}88), normal ({\em g}85) and maximum ({\em g}83) total kinetic
energy are shown in Fig.~\ref{fig:gradial88}. The vertical line at
$\log q \simeq-2.23$ marks the location of the maximum composition
gradient in the He/H transition region. It is evident that the {\em
  g}88 mode has lower amplitude below the He/H interface, so it is
trapped in the envelope. As the region below the He/H transition is
very dense, it has a high weight in the total kinetic energy, and
hence trapped modes have minimum values of the kinetic energy. We
recall that, for high-radial order {\em g}-modes in this model,
significant energy interchange is produced in the region below the
He/H transition, which is mostly damping (fig.~8, Paper~I), resulting
in trapped modes being less damped and having maximum growth rate
values. On the other hand, the {\em g}83 mode has highest amplitude in
the damping region, and it is referred to as a confined mode, which is
less likely to be excited, as is reflected in its minimum value of the
growth rate.

Previous adiabatic studies for sdBs \citep{charpinet00} postulated
qualitatively, based on the growth rate dependence of the kinetic
energy, $\eta\, \alpha \, W/E_{kin}$ (where $W$ is the work integral
defined as the total energy balance over one period of oscillation),
that when a mode had lower amplitude in the damping regions, the mode
would be less damped and would be more likely to be excited. Indeed,
we confirm this hypothesis for high-radial order sdO gravity modes in
our non-adiabatic study.

   \begin{figure*}
     \begin{tabular}{cc}
 \resizebox{0.45\linewidth}{6cm}{\includegraphics[angle=90]{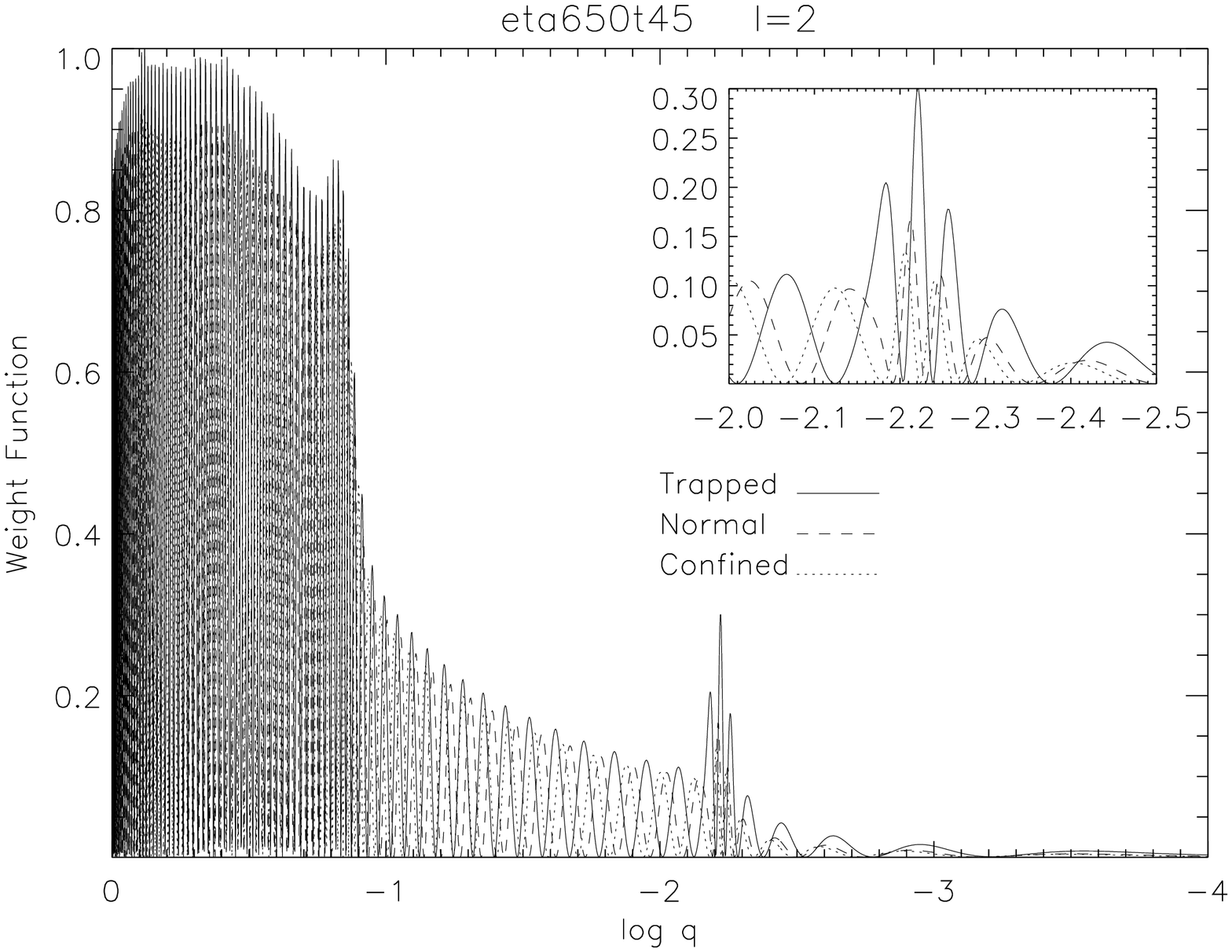}}
 &
 \resizebox{0.45\linewidth}{6cm}{\includegraphics[angle=90]{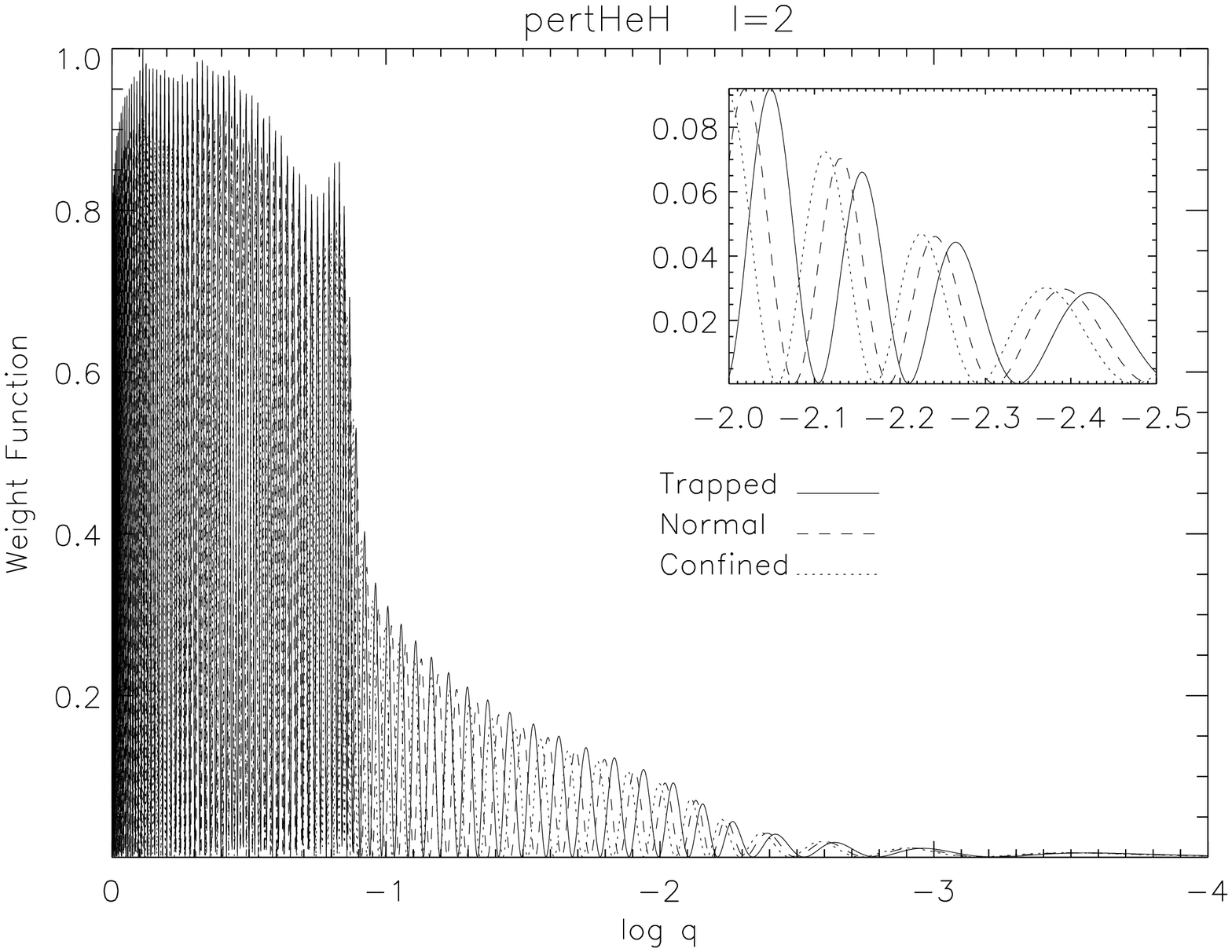}}
 \\
      \end{tabular}
   \caption{Weight function for {\em g}88, {\em g}85 and {\em g}83, a
     trapped, normal and confined mode, respectively for the original
     model (left) and modes with equivalent frequencies for the model
     without He/H chemical transition (right). Inset showing a blow-up
     of the He/H transition region.}
   \label{fig:grweif88}
   \end{figure*}

We also draw the attention to the fact that the radial eigenfunction
for the trapped mode has a node above the He/H transition, while the
confined mode has a node below it, and \ the normal mode has a node
just about on the interface. For the horizontal eigenfunction,
although there are nodes at similar distances above and below the He/H
transition, for the trapped (confined) mode, the closest node to the
transition is below (above) it, in good agreement which what is found
for sdBs and white dwarfs. However, the larger distances of the nodes
to the trapping region in our model translate into less efficiency of
the trapping/confinement mechanism.

In our model, the smaller differences in the kinetic energy between
trapped and confined modes comparing to sdB or white dwarf modes, are
due to the fact that the mode trapping interface (the He/H transition,
see below) is located much deeper in the envelope ($\log q \sim -2$)
than for sdBs ($\log q \sim -4$, in \citealt{charpinet00}) or white
dwarfs ($\log q \sim -10$, in \citealt{brassard92}). As a result, both
the trapped and confined region, weighted by the density, $\rho$ (see
Eq.~\ref{eq:ekin-mode}), contribute largely to the kinetic energy.

Also from \citet{charpinet00} and \citet{brassard92}, trapped modes in
sdBs and white dwarfs have sharp kinetic energy minima, while maxima,
corresponding to confined modes, are somewhat wider, comprising up to
three modes and indicating that confinement processes are not so
effective as mode trapping. In our model, both the minima and the
maxima in the kinetic energy have a certain width, revealing that both
the mode trapping and confinement processes are not so sharp as in
sdBs and white dwarfs. The explanation lies in the wider chemical
transition regions in the sdO model (see Fig.~\ref{fig:phys2epsn2},
right) in comparison to those for sdBs and white dwarfs (see
e.g. fig.~3 of \citealt{charpinet00} and fig.~1 of
\citealt{brassard92}).

   \begin{figure*}
     \begin{tabular}{cc}
 \resizebox{0.47\linewidth}{6cm}{\includegraphics{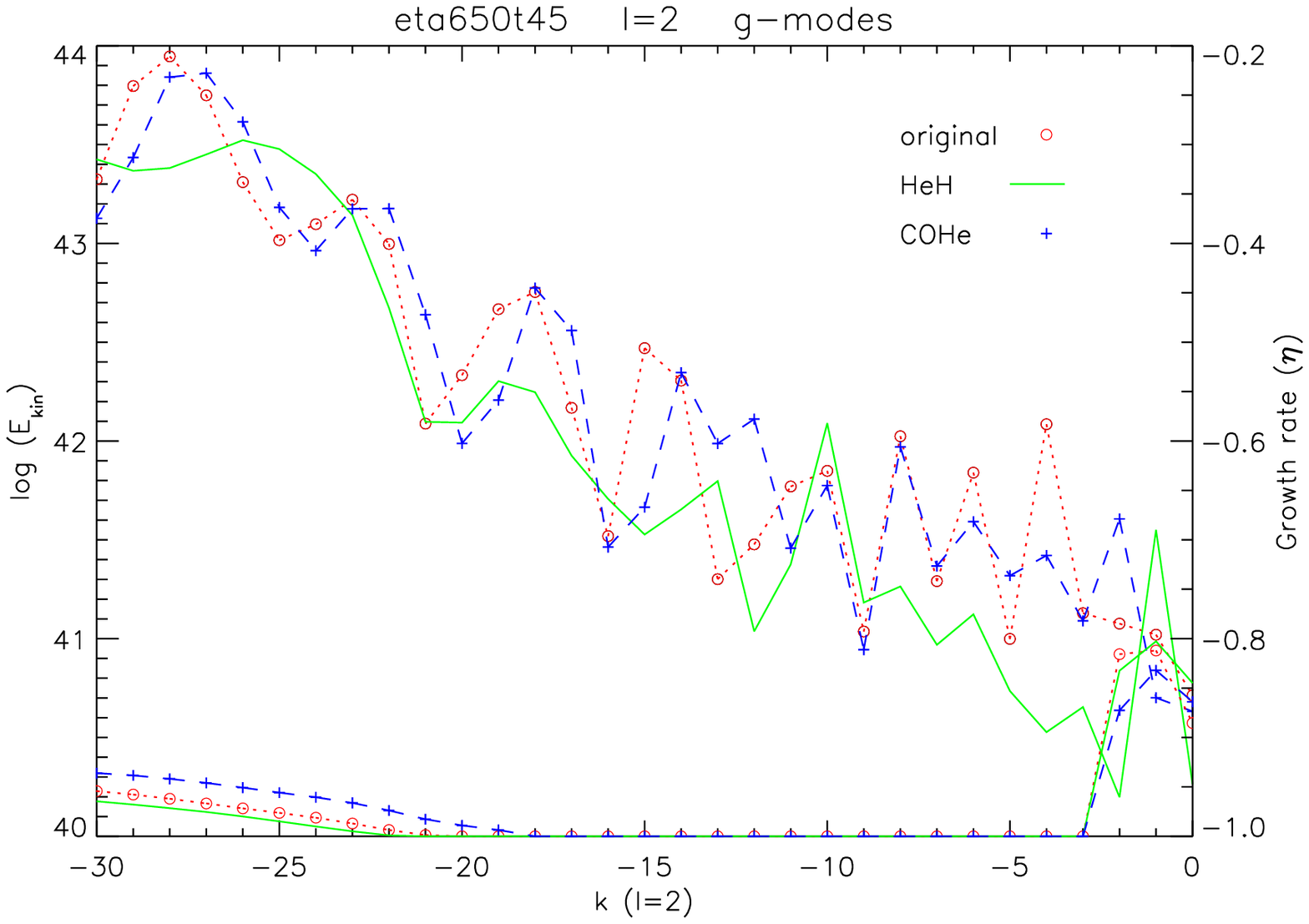}} &
 \resizebox{0.47\linewidth}{6cm}{\includegraphics{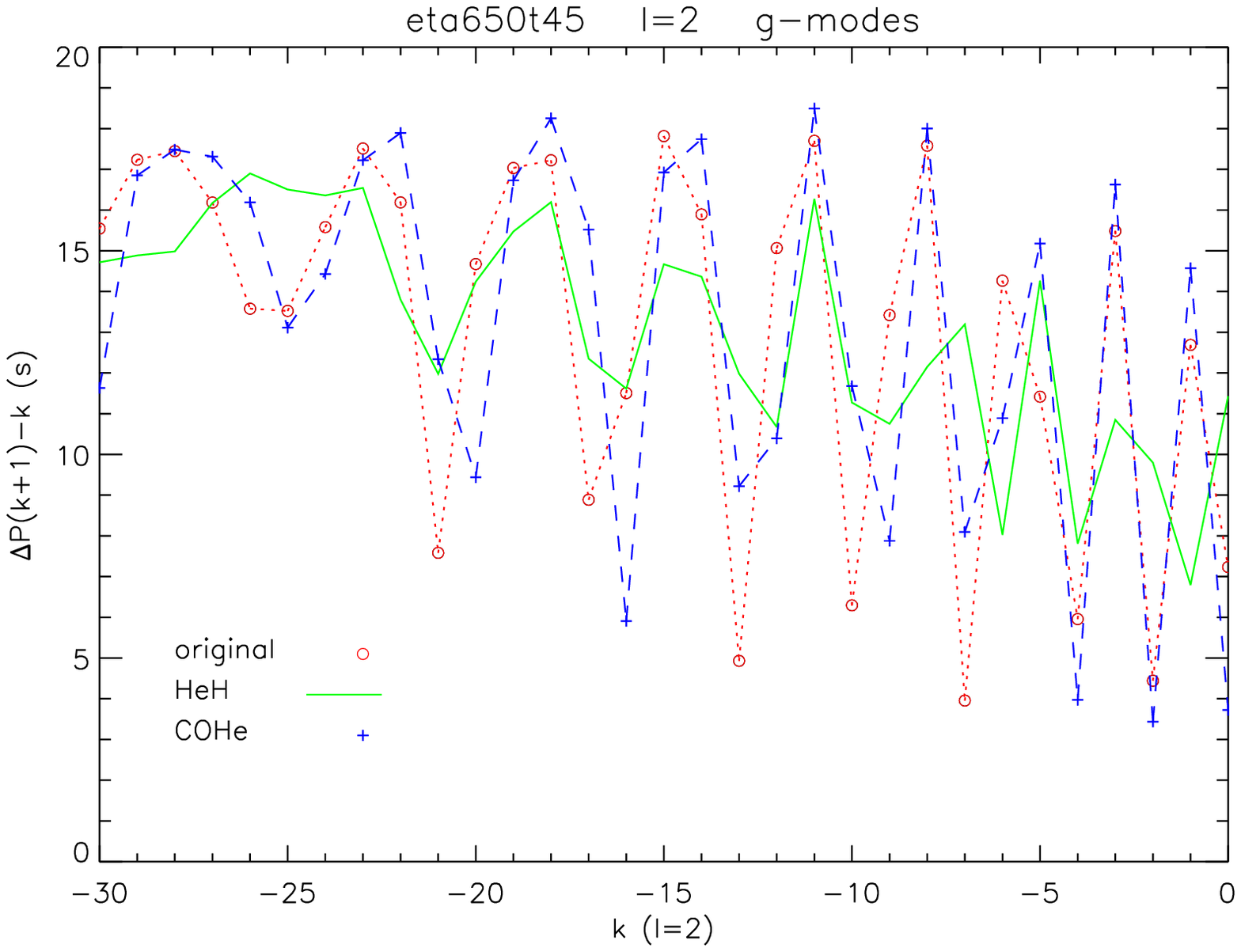}} \\
      \end{tabular}
   \caption{ Left: Logarithm of the kinetic energy vs. radial order
     for low-radial order {\em g}-modes of the original (upper circles
     and dotted line) and perturbed models without the He/H (upper
     solid line) and C-O/He (upper crosses and dashed line) transition
     regions. Lower symbols plot the corresponding growth rates whose
     values can be read on the right axis. Right: Period spacing for
     low-radial order {\em g}-modes with the same degree $\ell$ and
     consecutive radial orders for the original and perturbed He/H and
     C-O/He models (with same symbols as in the left pannel).}
   \label{fig:zoommiecin-deltap}
   \end{figure*}

   \begin{figure*}
     \begin{tabular}{cc}
 \resizebox{0.47\linewidth}{6cm}{\includegraphics{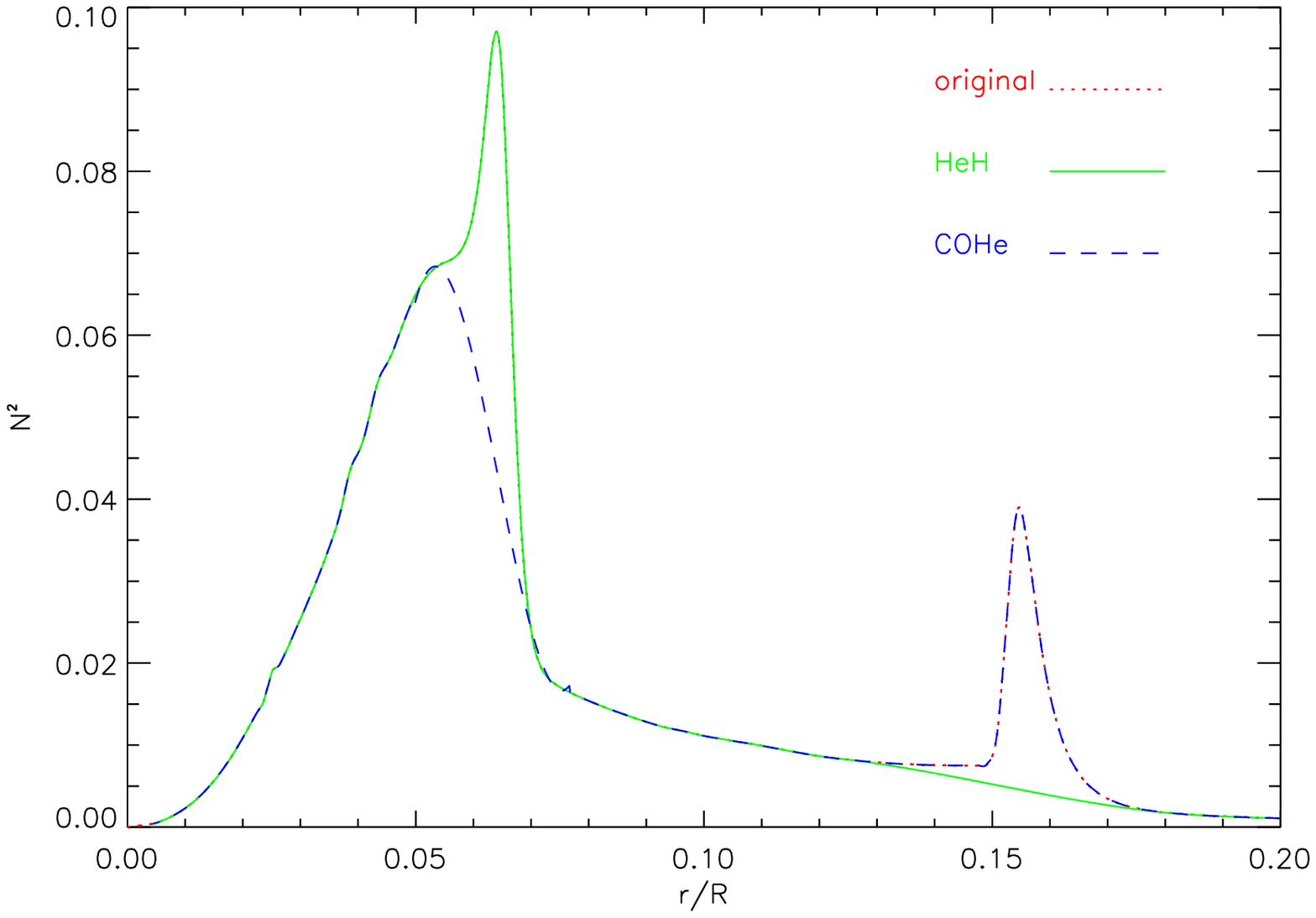}} &
 \resizebox{0.47\linewidth}{6cm}{\includegraphics[]{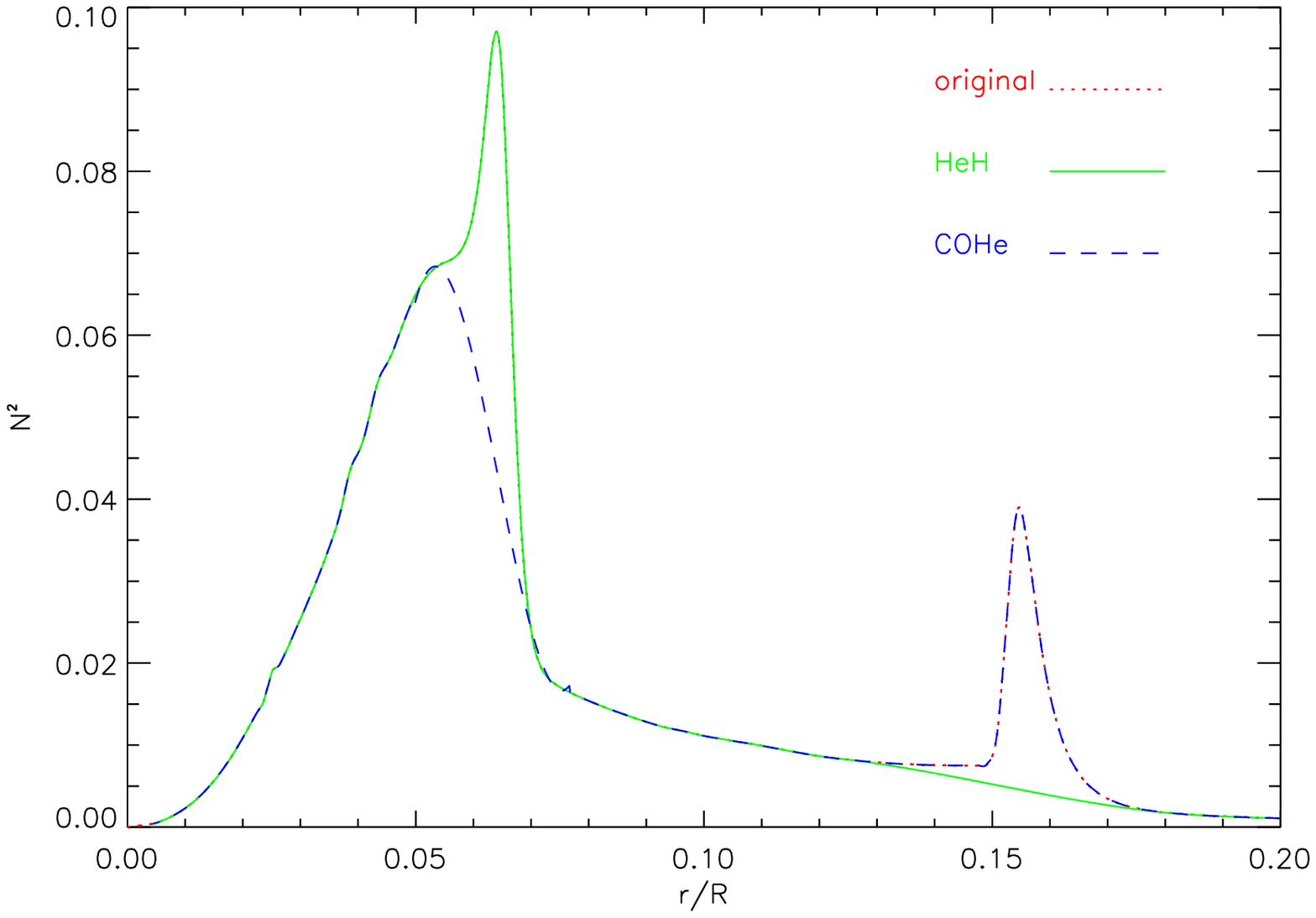}}
 \\
      \end{tabular}
   \caption{Original (dotted) Brunt-V\"ais\"al\"a frequency (left) and
     sound speed (right) and modifications cancelling out the He/H
     (solid) and sharp peak of the C-O/He (dashed) chemical transition
     regions.}
   \label{fig:n2pert}
   \end{figure*}

A clearer view of the local contribution of each region of the star to
the formation of a particular mode is given through the so-called
weight function, defined in \cite{charpinet00} as:

\begin{equation}
 F\left(\xi_r, P', \Phi'; r\right) = \left[\xi_r^2 N^2 +
   \frac{(P')^2}{\Gamma_1 P \rho} + \left(\frac{P'}{\Gamma_1 P} +
   \xi_r \frac{N^2}{g}\right) \right] \rho r^2
\label{eq:weifunction}
\end{equation}

where $P'$ and $\Phi'$ are the Eulerian perturbations of the pressure
and potential, respectively and $\Gamma_1$ is an adiabatic exponent.

A plot of the weight function for the same trapped ({\em g}88), normal
({\em g}85) and confined mode ({\em g}83) (Fig.~\ref{fig:grweif88}
left), each of them normalised individually, shows that the wide
C-O/He chemical transition region has an important contribution for
all of them, as expected for high-radial order {\em g}-modes, which
mainly propagate in the deep interior of the star. The main difference
in the weight function for the different modes is the higher
contribution of the He/H transition region to the formation of the
trapped mode, which also has a higher contribution of the outer
zones. This becomes even clearer when we eliminate in the model the
He/H chemical transition region (Fig.~\ref{fig:grweif88} right,
explanation below).

A blow up of Fig.~\ref{fig:allgkin} for the low-radial order {\em
  g}-modes ($k$ up to 30) is shown in the left panel of
Fig.~\ref{fig:zoommiecin-deltap} (circles and dotted lines), and the
period differences in this zone in the right panel. We still notice
the existence of certain modes alternating maximum and minimum values
of the kinetic energy, until about $k=10$. The period differences
between modes also show deviations with respect to the mean value, a
signature of mode trapping. However, contrary to high-radial order
{\em g}-modes, this trapping/confinement phenomenon, does not have any
influence on the ability to excite the modes, as they remain highly
stable. This is explained as, for {\em g}-modes of low-radial order,
the energy interchanged below the He/H transition is negligible
compared to the energy interchanged at the Z-bump region, located at
$-10.0 \lesssim \log q \lesssim -9.5$.  There, the energy contribution
for both trapped and confined modes is always damping (see fig.~8 in
Paper~I). So, even if the mechanical effect of the mode trapping is
maintained, at the frequencies in which low-radial order {\em g}-modes
occur, the significant energy interchange is produced in different
regions of the star, therefore the trapping has no influence in the
growth rate.

\subsection{Perturbation analysis}
We have investigated the effects of cancelling out the He/H and the
sharp peak of the C-O/He chemical transition regions of the
Brunt-V\"ais\"al\"a frequency and sound speed (Fig.~\ref{fig:n2pert})
in this sdO model on mode trapping and the tendency of modes to
instability\footnote{These new perturbed models will be referred to as
  pertHeH and pertCO, and their symbol codes in the plots are solid
  lines, and crosses and dashed lines respectively.}. To do this, we
described the perturbation of the mass, pressure, sound speed squared
($c^2$) and BV squared ($N^2$) as a function of the perturbation of
the density, gravity and adiabatic exponent. From these relations we
derived an expression for the perturbation of the density as a
function of the perturbations introduced for BV and the sound speed
($\delta N^2, \delta c^2$). Finally, we wrote the perturbation of the
dynamical variables of the model as a function of the already
calculated perturbations (a complete description of the mathematical
formulation is given in Appendix A). These new perturbed models
underwent again the non-adiabatic analysis.

 \begin{figure}
   \includegraphics[width=9cm]{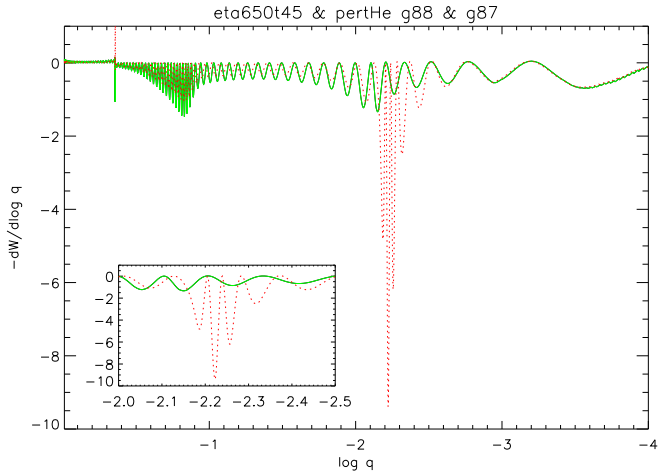}
   \caption{Differential work for {\em g}88 mode of the original (dotted)
     model and {\em g}87 mode of the pertHeH (solid) model. The inset
     zooms in on the He/H chemical transition. See text for details.}
   \label{fig:dwzoom}
   \end{figure}

 \begin{figure}
   \includegraphics[width=8.5cm]{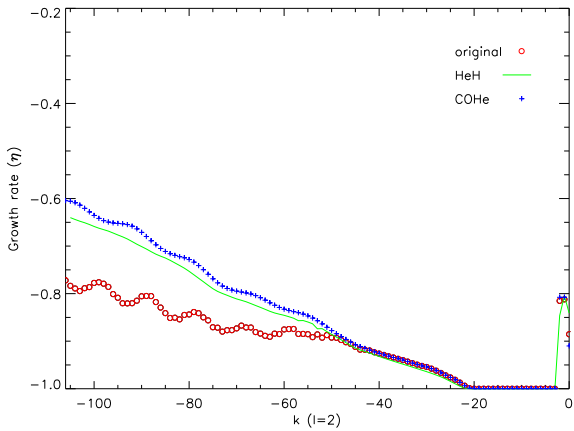}
   \caption{Growth rates of the original model (circles), the same model
     with the differential work integrated only over $\log q \leq
     -2.24$ (crosses) and the pertHeH model (solid line).}
   \label{fig:cmpgrou}
   \end{figure}

 \begin{figure}
   \includegraphics[width=9cm]{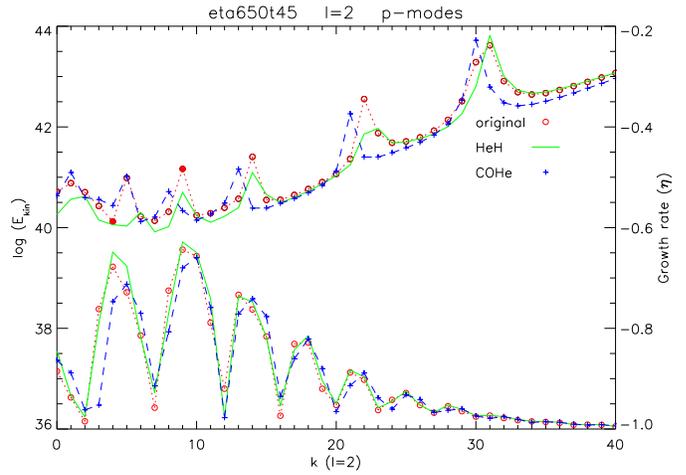}
   \caption{Logarithm of the total kinetic energy vs. radial order for
     {\em p}-modes of the original (upper circle and dotted line) and
     perturbed models without the He/H (upper solid line) and C-O/He
     (upper crosses and dashed line) transition regions. Lower symbols
     plot the corresponding growth rates whose values can be read on
     the right axis. Filled circles mark {\em p}4 and {\em p}9 modes
     showing maximum growth rate values. See Section~5 for more
     details.}
   \label{fig:allpkin}
   \end{figure}

   \begin{figure}
   \includegraphics[width=9cm]{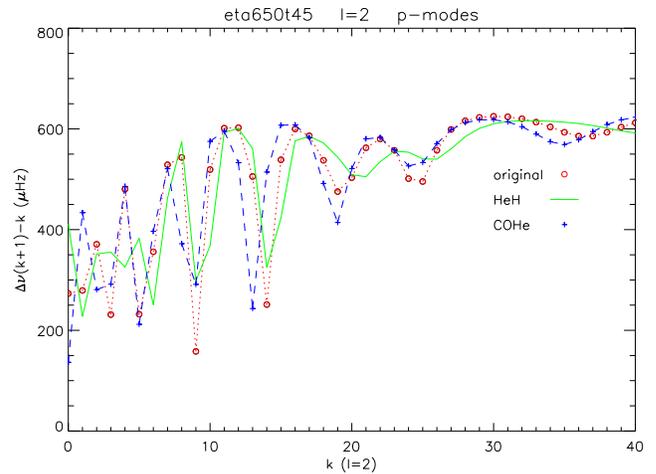}
   \caption{Large frequency spacing vs. radial order for {\em p}-modes
     with the same degree $\ell$ and consecutive radial orders for the
     original, and perturbed models.}
   \label{fig:allpdeltanuk}
   \end{figure}

\subsubsection{Cancelling out the He/H transition}
When we cancel out the He/H chemical transition region, we get a
smooth kinetic energy for all the modes (Fig.~\ref{fig:allgkin}, upper
solid line). Trapped modes vanish, revealing that the He/H transition
is the main factor responsible for their occurrence. The plot of the
weight function (Fig.~\ref{fig:grweif88}, right) reveals that the He/H
region, once cancelled, no longer has any significant influence on
mode formation. We also note that the growth rate no longer oscillates
with frequency (Fig.~\ref{fig:allgkin}, lower solid line), as expected
from the uniform kinetic energy; and we also get a uniform period
separation (Fig.~\ref{fig:allgdeltap}, solid line). However, perturbed
modes equivalent to trapped modes in the original model achieved
higher values of the growth rate, while their kinetic energy was
higher, which seems to be at odds with the growth rate dependence on
the kinetic energy.

The local minimum in the damping energy for model 8 at $\log q=-2.23$
(Fig.~\ref{fig:dwzoom}, dotted line), which is absent in model pertHeH
(Fig.~\ref{fig:dwzoom}, solid line), is responsible for the overall
lower values of the growth rate in the original model. Thus, when we
calculate the work function for the original model integrating between
$\log q \leq -2.24$ and the surface, we obtain similar growth rate
values as those of the pertHeH model (Fig.~\ref{fig:cmpgrou}).

In the case of low radial order {\em g}-modes
(Fig.~\ref{fig:zoommiecin-deltap}, solid lines) some subtle mode
trapping seems still to be present. This may be due to remnant mode
trapping effects caused by the C-O/He transition, in the same way as
described below for low radial order {\em p}-modes.

   \begin{figure*}
     \begin{tabular}{cc}
 \resizebox{0.45\linewidth}{6cm}{\includegraphics[angle=90]{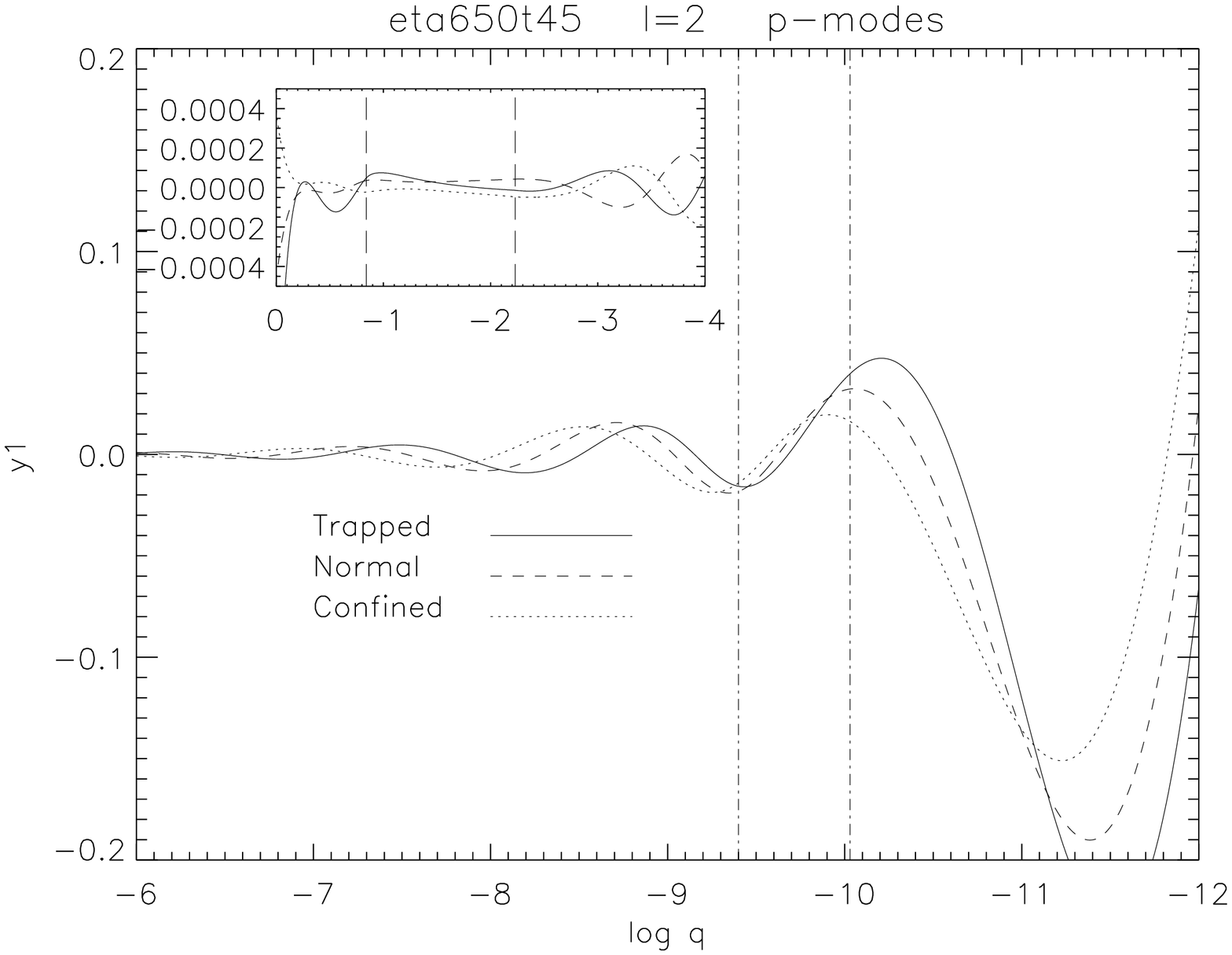}}
 &
 \resizebox{0.45\linewidth}{6cm}{\includegraphics[angle=90]{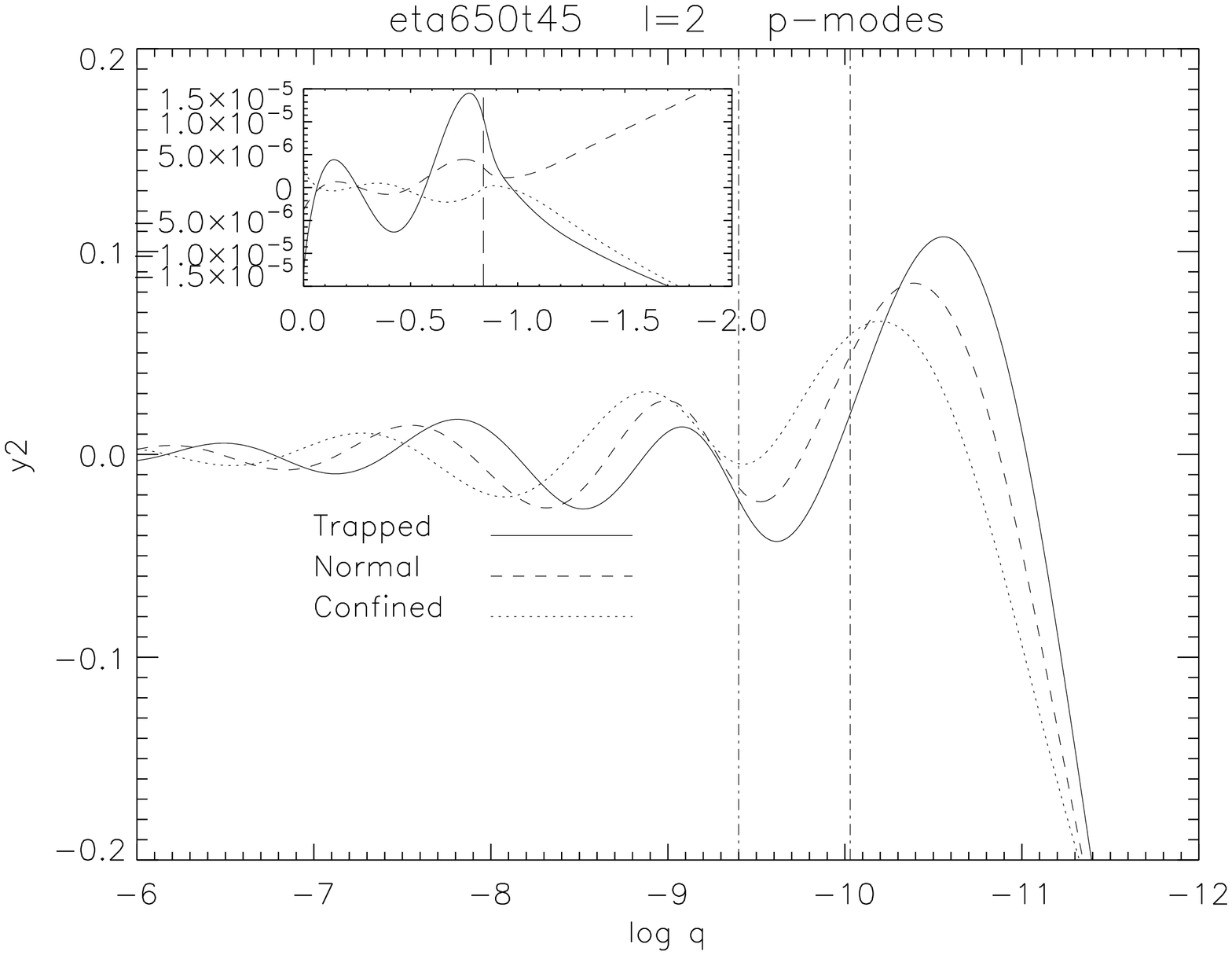}}
 \\
      \end{tabular}
   \caption{Radial (left) and horizontal (right) displacement
     eigenfunctions for a trapped, normal and confined {\em p}-mode of
     the original model:{\em p}9, {\em p}8 and {\em p}7,
     respectively. Dashed-dotted vertical lines delimit the convective
     region where maximum opacity takes place. Dashed vertical lines
     in the insets mark the maxima of the C-O/He (the deepest) and
     He/H transition regions.}
   \label{fig:p9p8p7y1y2}
   \end{figure*}

\subsubsection{Cancelling out the C-O/He transition}
When we cancel out the sharp peak of the C-O/He chemical transition,
keeping the He/H peak, we still obtain the original pattern of trapped
modes, with total kinetic energy, growth rate and period differences
similar to those of the original model (Fig.~\ref{fig:allgkin},
Fig.~\ref{fig:allgdeltap} crosses and dashed lines). Thus, this
chemical transition plays no role in high radial order {\em g}-mode
trapping. However, for lower order modes
(Fig.~\ref{fig:zoommiecin-deltap}, left panel, crosses and dashed
lines), it may have some influence, which is revealed in certain
alterations to the total kinetic energy of each mode.

\section{Pressure modes}
The logarithm of the total kinetic energy, and corresponding growth
rates, for the {\em p}-mode spectrum are shown in
Fig.~\ref{fig:allpkin}. The depicted radial orders correspond to
frequencies ranging from about 6 to 25~mHz. There are certain modes
which show values higher than the mean kinetic energy in the original
and perturbed models. We investigate if this could also be an effect
of mode trapping.

Following again the asymptotic theory of nonradial oscillations
(\citealp{tassoul80}; \citealp{smeyers95}; \citealp{smeyers07}) {\em
  p}-modes of the same degree $\ell$ and consecutive radial order $k$,
will have constant frequency separation, $\Delta \nu$, given by :

\begin{equation}
\Delta \nu = \nu_{k+1,l} - \nu_{k,l}= \frac{1}{2} \left[ \int_0^R
  \left(\frac{\rho}{\Gamma_1 p}\right)^{1/2} dr \right]^{-1}
\label{eq:separeixon}               
\end{equation}

Thus, the mode trapping signature for {\em p}-modes would show up in
the deviations in the large frequency spacing for the original and the
perturbed He/H model (Fig.~\ref{fig:allpdeltanuk}).

   \begin{figure*}
     \begin{tabular}{cc}
 \resizebox{0.45\linewidth}{6cm}{\includegraphics[angle=90]{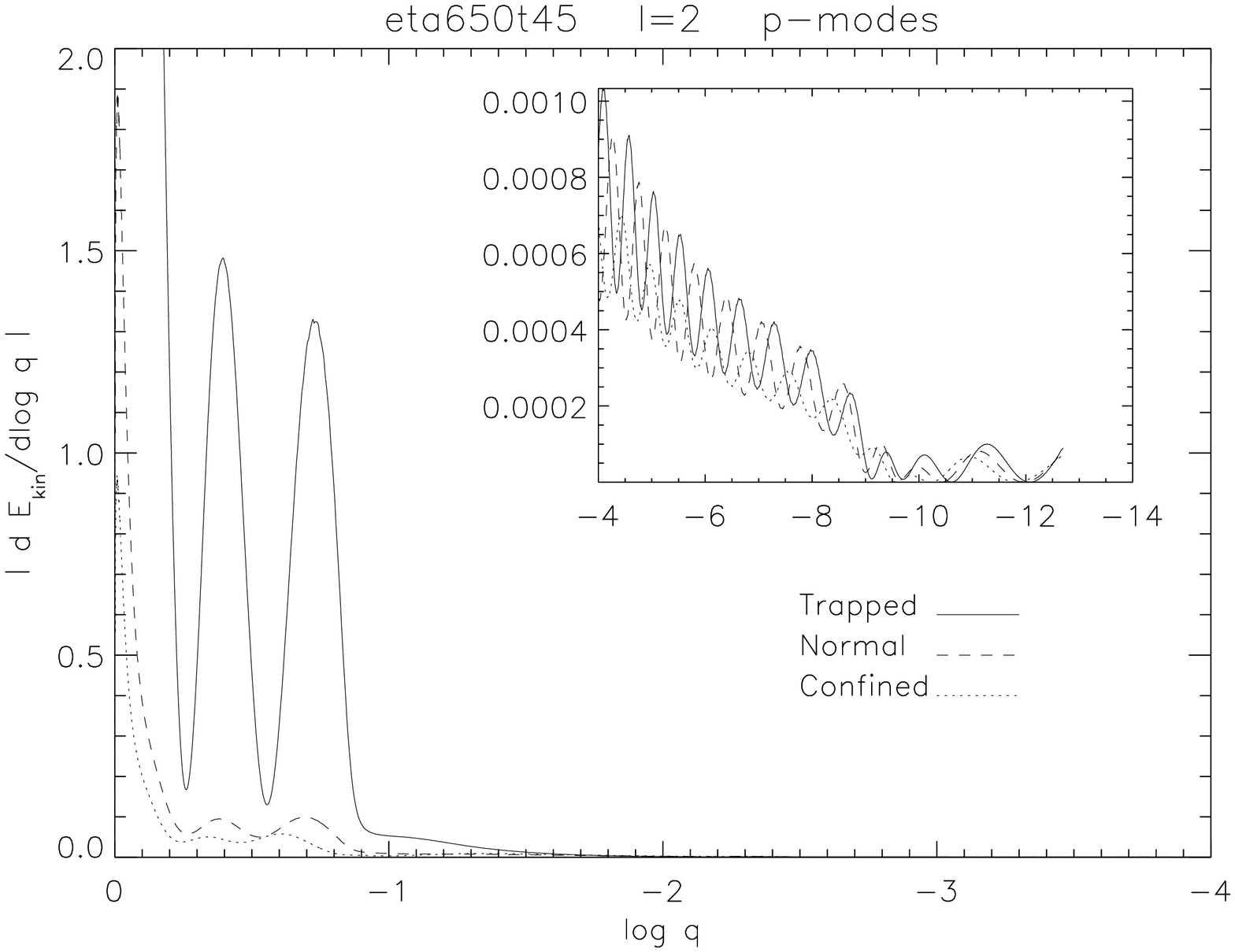}}
 & \resizebox{0.45\linewidth}{6cm}{\includegraphics{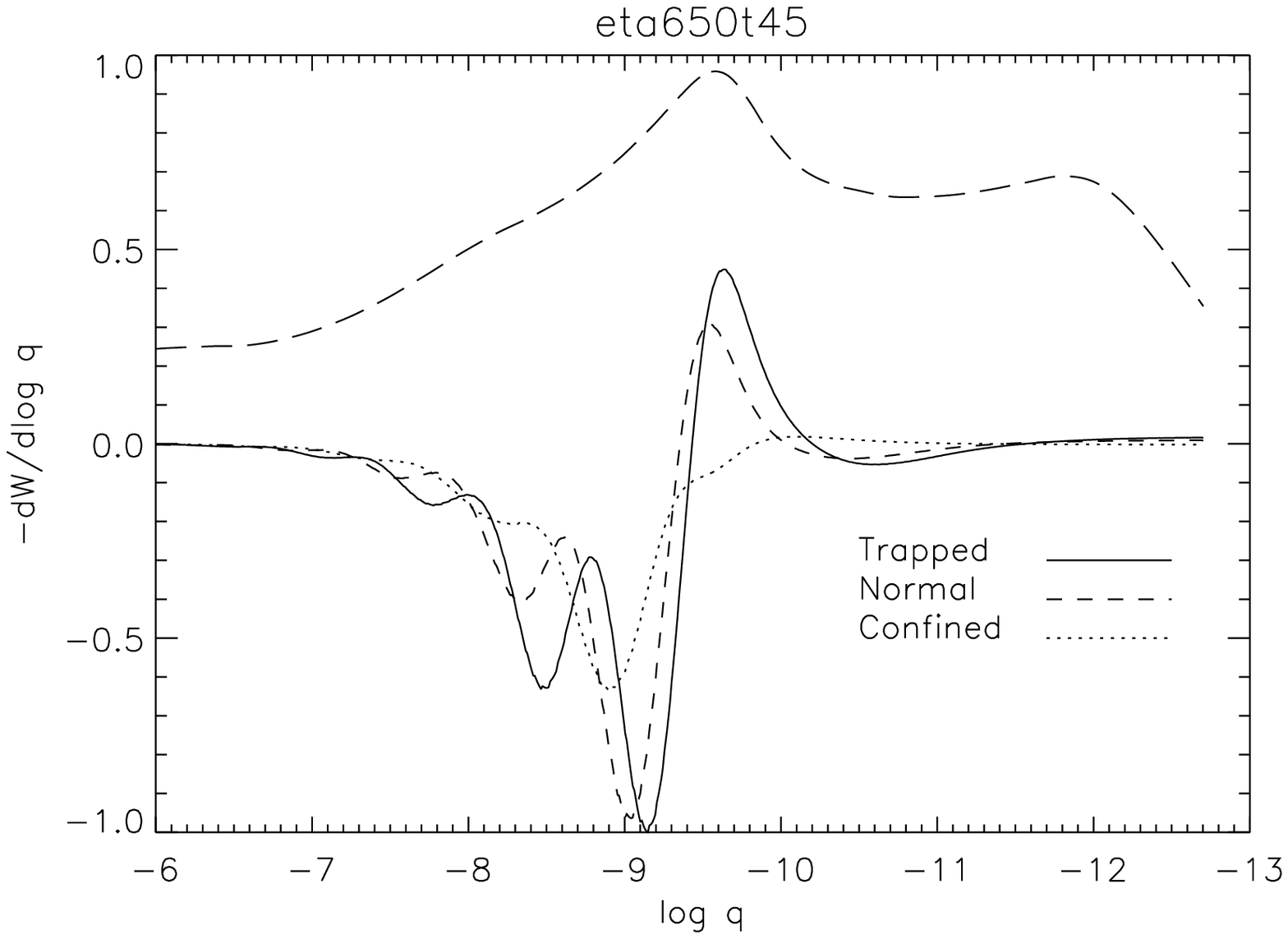}} \\
      \end{tabular}
   \caption{Left: Local kinetic energy for trapped, normal and
     confined modes: {\em p}9, {\em p}8, {\em p}7,
     respectively. Right: Differential work and opacity (long dashed
     line) for the same modes. The maximum of the Z-bump is located at
     $\log q \simeq -9.5$.}
   \label{fig:indecinet-trabajop9p8p7}
   \end{figure*}

We have plotted the radial and horizontal displacement eigenfunctions
for the original model {\em p}-modes with maximum ({\em p}9),
intermediate ({\em p}8) and minimum ({\em p}7) total kinetic energy
(Fig.~\ref{fig:p9p8p7y1y2}). We found that all modes show higher
amplitude in the envelope, as expected for {\em p}-modes. We found
that the {\em p}9 mode shows the highest amplitude in the outer
envelope for the 3 modes under comparison. However, ultimately what
accounts for the higher total kinetic energy is the higher amplitude
of the eigenfunctions in the innermost layers of the star, where the
mode is trapped due to the pinching effect of the C-O/He interface on
the eigenfunctions. This can be seen in
Fig.~\ref{fig:indecinet-trabajop9p8p7} (left) that shows the local
contribution to the kinetic energy of the trapped, normal and confined
modes, {\em p}9, {\em p}8, {\em p}7, respectively, normalised to the
maximum of {\em p7}. Although not shown for the sake of clarity, the
maximum kinetic energy value for the {\em p}9 trapped mode reaches one
order of magnitude higher.

The plot of the differential work for a trapped, normal and confined
mode (Fig.~\ref{fig:indecinet-trabajop9p8p7}, right) shows that
significant energy interchange is only produced at the Z-bump; and
there, driving is higher for the trapped mode.  As it is the case for
low-radial order {\em g}-modes, a maximum in kinetic energy does not
translate into a minimum growth rate value for the mode. This is
explained by the maximum kinetic energy being due to a higher
amplitude of the eigenfunctions in the region $\log q \gtrsim -0.8$,
which for {\em p}-modes does not have significant influence on
driving. Neither do the higher amplitudes of the eigenfunctions of the
trapped mode at the driving region lead to a maximum growth rate
value, as consecutive higher radial order modes have even higher
amplitudes.  We conclude that the influence of the work function
prevails over the influence of the kinetic energy in the computation
of the growth rate.

Pressure modes with higher radial orders, not shown in
Fig.~\ref{fig:allpkin}, display kinetic energy monotonically
increasing, an effect of the nodes accumulating in the surface limit
imposed by the boundary conditions, as it is well described in
\citet{charpinet00}. The trapping effect caused by the C-O/He
transition is no longer produced, as is expected for high-radial order
pure {\em p}-modes propagating only in the external layers of the
star.

\subsection{Perturbation analysis}
When we cancel the He/H, or the sharp peak of the C-O/He chemical
transitions in the Brunt-V\"ais\"al\"a frequency, and also the
corresponding bumps in the sound speed at these locations, we still
find a kinetic energy pattern, and an oscillating profile of the
growth rate, both similar to those of the original model
(Fig.~\ref{fig:allpkin}, solid and dashed lines,
respectively). Therefore, neither the He/H, nor the C-O/He chemical
transition, nor the sound speed at these locations, seem to have any
significant influence on the kinetic energy or on the tendency to
driving of the {\em p}-modes.

This result was the expected for the growth rate, for reasons given
above, but not for the kinetic energy, as we presumed the modification
of the C-O/He transition would show in the kinetic energy profile, due
to changes in the amplitude of the eigenfunctions. However,
eigenfunction profiles similar to those of Fig.~\ref{fig:p9p8p7y1y2}
are retained for trapped, normal and confined modes of the perturbed
models. The reason may be due to that, in fact, the C-O/He transition
still exists (see Fig.~\ref{fig:n2pert}, left) as we only cancelled
out its steep peak, giving a softer profile.

\section{Summary}
We conclude for the {\em g}-mode spectrum that:
\begin{itemize}
\item High-radial order {\em g}-modes are trapped in the envelope due
  to the He/H chemical transition. This provides a weak selection
  mechanism, in the way that trapped modes oscillate with lower
  amplitudes in the innermost damping region. Thus, trapped modes show
  lower total kinetic energy and higher values of the growth rate.

\item Low to intermediate radial order {\em g}-modes (mixed modes) are
  also trapped in the envelope by the He/H transition. However, this
  has no significant effects on the driving, as the innermost damping
  region, at the frequencies at which low {\em g}-modes occur, has no
  significant weight in the energy interchanged by the mode with its
  surroundings.
 
\item Cancelling out the He/H transition region cancels the mode
  trapping effects of high order gravity modes, and the growth rate
  loses its oscillatory behaviour with frequency, as
  expected. However, modes with local maxima in the kinetic energy and
  similar interchanged energy at the driving region than their
  non-perturbed counterparts, achieve higher growth rate values. This
  effect is due to the presence of a small but strong damping region
  in the original model at the location of He/H chemical transition,
  which has an overall damping effect in the high-radial order {\em
    g}-modes of the original model. For low radial order gravity modes
  some remnant trapping still exists which may be attributed to the
  C-O/He transition.

\item When we cancel out the sharp peak of the C-O/He transition
  region, keeping the He/H peak, mode trapping is still present for
  high order {\em g}-modes, and also the oscillating behaviour of the
  growth rate. Thus, the C-O/He transition region has no significant
  role in their trapping. However, it may play a subtle role in low
  order {\em g}-mode trapping (seen in certain alterations to the
  total kinetic energy), due to the persistence of the
  transition. Trapping though, has no effect in their driving.

We also have to bear in mind that although the Brunt-V\"ais\"al\"a
frequency is expected to have a higher weight in the mode trapping,
the contribution of the sound speed has not been disentangled in the
analysis and may be responsible in part for certain aspects of
trapping.
\end{itemize}

We conclude for the {\em p}-mode spectrum that:
\begin{itemize}
\item Low to intermediate radial order {\em p}-modes (mixed modes) are
  found to be trapped by the C/O-He transition in the innermost
  region. However, this has no significant effects on the driving as
  the maximum energy interchanged is produced in the envelope of the
  star.
\item Cancelling out the He/H or the sharp peak of the C-O/He
  transition has no significant effect neither in the mode trapping,
  nor in the growth rate profile, which remain essentially
  unaltered. This is interpreted as trapping effects produced by the
  remaining piece of the C-O/He transition.
\end{itemize}

\section{Discussion and Conclusions}
A sdO model presented an oscillating profile of the growth rate with
frequency, which lead us to analyse it in more detail: it was found
that gravity modes suffered mode trapping effects caused by the He/H
chemical transition region which made high-order {\em g}-trapped modes
more likely to be driven. Mode trapping was also found for low-radial
order {\em p}-modes, caused by the deep C-O/He chemical
interface. However, no obvious correlation could be made with its
possible non-adiabatic effects.

We sought to investigate the non-adiabatic effects of mode trapping
through the influence of the kinetic energy on the growth rate
behaviour. Up to now, the existing adiabatic studies for sdBs
\citep{charpinet00} qualitatively proposed that, as a trapped mode had
lower amplitude in the damping regions, the mode would be less damped
and would be more likely to be excited. Indeed, this is what we found
in our non-adiabatic study for trapped in the envelope high-radial
order {\em g}-modes: they oscillate with lower amplitudes below the
He/H transition, a damping region, hence they have higher values of
the growth rate than its confined counterparts. In this way, mode
trapping acts a weak selection mechanism rendering trapped modes more
likely to be driven.

Meanwhile, low- to intermediate-radial order {\em p}-modes, which show
a similar behaviour of the growth rate, were found to be trapped by
the C-O/He transition. However, as the significant energy interchange
at {\em p}-mode frequencies is produced much higher in the envelope,
mode trapping under the C-O/He interface does not play a significant
role in the potential to destabilise modes.

Therefore, we conclude that mode trapping is a potential aid for
destabilizing modes when (1) modes are trapped in such a way that they
have lower amplitude in a damping region and (2) this region plays
a significant role in the energy interchanged by the mode with its
surroundings.

Finally, a tentative exercise of a theoretical mode identification of
the radial order of the observed modes of J1600+0748 was done in the
terms of asymptotic analysis, based on the similar mean frequency
spacing of our model 8 and the observed frequencies. We associated the
latter with those showing highest growth rate in model 8, and gaps in
the observed frequencies to modes with minimum growth rate. This
resulted in the observed modes associated with high radial-order {\em
  p}-modes, in contradiction with low-order low-degree mode
identification by Fontaine et al. (2008). However, we consider this
possible application of mode trapping to explain the observed sdO
pulsation spectra, a path that is worth exploring. For this exercise
to be fruitful, proper theoretical models reproducing the observed
physical parameters of J1600+0748 should be used.

\section*{Acknowledgments}
CRL acknowledges an {\em \'Angeles Alvari\~ no} contract of the
regional government {\em Xunta de Galicia}. This research was also
supported by the Spanish Ministry of Science and Technology under
project ESP2004--03855--C03--01 and by the \emph{Junta de Andaluc\'ia}
and the {\em Direcci\'on General de Investigaci\'on (DGI)} under
project AYA2000-1559. AM acknowledges financial support from a {\em
  Juan de la Cierva} contract of the Spanish Ministry of Education and
Science. RO is supported by the Research Council of Leuven University
through grant GOA/2003/04. The research of JM is supported in part by
a grant from the Mount Cuba Astronomical Foundation.

\section*{Appendix A}
We describe here the modifications imposed on the dynamical variables
of the model through perturbations of Brunt-V\"ais\"al\"a frequency
($\delta N^2$) and the sound speed ($\delta c^2$). The strategy that
we follow is first to describe the perturbation of physical variables
of mass, pressure, $N^2$ and $c^2$ as a function of the perturbation
of the density ($\delta \rho$), gravity ($\delta g$) and adiabatic
exponent ($\delta \Gamma_1$). Then, we derive an expression for the
perturbation of the density as a function of $\delta N^2$, $\delta
c^2$. Finally, we will describe the perturbation of the dynamical
variables of the model as a function of the already calculated
perturbations.

We begin perturbing the equations of stellar structure:

\begin{equation}
\frac{dm}{dr} = 4 \pi \rho r^2 \hspace{0.3cm}, \hspace{0.3cm} \delta m
= \int_0^R 4 \pi r^2 \delta \rho dr
\label{eq:deltam}
\end{equation}

\begin{equation}
g = \frac{Gm}{r^2} \hspace{0.3cm},\hspace{0.3cm} \frac{\delta g}{g} =
\frac{\delta m}{m}
\label{eq:deltag}
\end{equation}

\begin{equation}
\frac{dp}{dr} = - \rho g \hspace{0.3cm},\hspace{0.3cm} \frac{d \delta
  p}{dr} = - \rho g \left(\frac{\delta \rho}{\rho} + \frac{\delta
  g}{g} \right) \hspace{0.3cm},
\end{equation}

\begin{equation}
\delta p = \int_0^R \rho g \left( \frac{\delta \rho}{\rho} +
\frac{\delta g}{g} \right) \hspace{0.3cm},\hspace{0.3cm}
\end{equation}

We perturb the sound speed:

\begin{equation}
c^2 = \frac{\Gamma_1 p}{\rho} \hspace{0.3cm},\hspace{0.3cm} \delta c^2
= c^2 \left(\frac{\delta \Gamma_1}{\Gamma_1} + \frac{\delta p}{p} -
\frac{\delta \rho}{\rho}\right)
\label{eq:c2}
\end{equation}

And Brunt-V\"ais\"al\"a frequency:

\begin{eqnarray}
N^2 & =& g \left(\frac{1}{\Gamma_1}\frac{d \ln p}{dr} - \frac{d \ln
  \rho}{dr}\right) = \frac{g}{\Gamma_1 p}\frac{dp}{dr} -
\frac{g}{\rho}\frac{d \rho}{dr} = \nonumber \\ N^2 & =&
-\frac{g^2}{c^2} - \frac{g}{\rho} \frac{d \rho}{dr}
\label{eq:n2}
\end{eqnarray}

\begin{equation}
\delta N^2 = \frac{\delta g}{g} N^2 -\frac{g \delta
  \Gamma_1}{\Gamma_1^2} \frac{1}{p} \frac{dp}{dr} + \frac{g}{\Gamma_1}
\frac{d}{dr}\left(\frac{\delta p}{p}\right) - g
\frac{d}{dr}\left(\frac{\delta \rho}{\rho}\right)
\label{eq:deltan2}
\end{equation}

Using the equations we have derived, we operate each term in the
equation above as:

\begin{displaymath}
-\frac{g \delta \Gamma_1}{\Gamma_1^2} \frac{1}{p} \frac{dp}{dr} =
\frac{g^2}{c^2} \left(\frac{\delta c^2}{c^2} - \frac{\delta p}{p} +
\frac{\delta \rho}{\rho} \right)
\end{displaymath}

\begin{displaymath}
\frac{g}{\Gamma_1} \frac{d}{dr} \left(\frac{\delta p}{p}\right) =
\frac{g}{\Gamma_1}\left(-\frac{1}{p^2}\frac{dp}{dr} \delta p +
\frac{1}{p} \frac{d \delta p}{dr}\right) = \frac{g^2}{c^2}
\left(\frac{\delta p}{p} - \frac{\delta \rho}{\rho} - \frac{\delta
  g}{g}\right)
\end{displaymath}

\begin{displaymath}
- g \frac{d}{dr}\left(\frac{\delta \rho}{\rho}\right) =
\frac{g}{\rho^2} \frac{d \rho}{dr} \delta \rho -\frac{g}{\rho}\frac{d
  \delta \rho}{dr}
\end{displaymath}

Replacing these three terms in Eq.~\ref{eq:deltan2} we obtain:

\begin{equation}
\delta N^2 = \left(N^2 - \frac{g^2}{c^2}\right) \frac{\delta g}{g} +
\frac{g^2 \delta c^2}{c^2 c^2} + \frac{g}{\rho} \frac{d \rho}{dr}
\frac{\delta \rho}{\rho} - \frac{g}{\rho}\frac{d \delta \rho}{dr}
\end{equation}

Replacing $N^2$ by Eq.~\ref{eq:n2} we obtain the integro-differential
equation in $\delta \rho$:

\begin{equation}
\frac{d \delta \rho}{dr} - \frac{d \rho}{dr} \frac{1}{\rho} \delta
\rho + \frac{\rho}{g} \left(\delta N^2 - \frac{g^2 \delta c^2}{c^2
  c^2}\right) + \frac{\rho}{g} \left(\frac{2 g^2}{c^2} +
\frac{g}{\rho} \frac{d \rho}{dr}\right) \frac{\delta g}{g} = 0
\end{equation}

Substituting Eq.~\ref{eq:n2} in the last term, replacing $\delta g$ by
Eq.~\ref{eq:deltag} and using Eq.~\ref{eq:deltam} we find:

\begin{equation}
\frac{\rho}{g} \left(\frac{2 g^2}{c^2} + \frac{g}{\rho} \frac{d
  \rho}{dr}\right) \frac{\delta g}{g} = \left(\frac{g^2}{c^2} - N^2
\right) \frac{\rho}{g m} \int_0^R 4 \pi r^2 \delta \rho dr
\end{equation}

Numerical tests indicate that this last term is negligible. Thus, we
are left with a linear first order differential equation of the type
$y'+a(x)y+b(x)=0$ with solution: $y=\frac{\int c(x) b(x) dx
  +C}{c(x)}$, with $c(x)=e^{\int a(x) dx}$. Thus, solving for $\delta
\rho$ we get:

\begin{equation}
\frac{\delta \rho}{\rho} = - \int_0^R \frac{1}{g} \left(\delta N^2 -
\frac{g^2 \delta c^2}{c^2 c^2}\right) dr
\end{equation}

In this way we have the perturbed density $\delta \rho$ as a function
of the perturbation of Brunt-V\"ais\"al\"a $\delta N^2$ and the sound
speed $\delta c^2$.

Finally, we perturb five independent dynamical variables of the twenty
four input variables which describe our equilibrium models. These
input variables are the same which describe the equilibrium models in
{\scriptsize ADIPLS} adiabatic code \citep{jcd08} and are convenient
when the equations are formulated as in \citet{dziembowski71}, as it
is our case:

\begin{equation}
c_1 = \frac{M}{m}
\left(\frac{r}{R}\right)^3 \hspace{0.3cm},\hspace{0.3cm} \delta c_1 =
- c_1 \frac{\delta g}{g}
\end{equation}

\begin{equation}
V_g = -\frac{1}{\Gamma_1} \frac{d \ln p}{d \ln r} = \frac{\rho g
  r}{\Gamma_1 p} = \frac {G m}{c^2 r} \hspace{0.3cm},\hspace{0.3cm}
\delta V_g = V_g \left(\frac{\delta g}{g} - \frac{\delta
  c^2}{c^2}\right)
\end{equation}

From Eq.~\ref{eq:c2} we have the perturbation of the adiabatic
exponent:

\begin{equation}
\delta \Gamma_1 = \Gamma_1 \left(\frac{\delta c^2}{c^2} - \frac{\delta
  p}{p} + \frac{\delta \rho}{\rho}\right)
\end{equation}

\begin{equation}
A^* = r \frac{N^2}{g} \hspace{0.3cm},\hspace{0.3cm} \delta A^* = A^*
\left(\frac{\delta N^2}{N^2} - \frac{\delta g}{g}\right)
\end{equation}

And the perturbation of the homology invariant:

\begin{equation}
U = \frac{4 \pi r^3 \rho }{m} \hspace{0.3cm},\hspace{0.3cm} \delta U =
U \left(\frac{\delta \rho}{\rho} - \frac{\delta g}{g}\right)
\end{equation}


\begin{thebibliography}{}

 \bibitem[\protect\citeauthoryear{Brassard et al.}{1991}]{brassard91}
   Brassard P., Fontaine G., Wesemael F., Kawaler S. D., Tassoul M.,
   1991, \apj, 367, 601

 \bibitem[\protect\citeauthoryear{Brassard et al.}{1992}]{brassard92}
   Brassard P., Fontaine G., Wesemael F., Hansen C. J., 1992, \apjs,
   80, 369

 \bibitem[\protect\citeauthoryear{Charpinet et
     al.}{2000}]{charpinet00} Charpinet S., Fontaine G., Brassard P.,
   Dorman, B., 2000, \apjs, 131, 223

 \bibitem[\protect\citeauthoryear{Charpinet, Fontaine \&
     Brassard}{Charpinet et al.}{2009}]{charpinet09} Charpinet S.,
   Fontaine G., Brassard P., 2009, A\&A, 493, 595

 \bibitem[\protect\citeauthoryear{Christensen-Dalsgaard}{2008}]{jcd08}
   Christensen-Dalsgaard J., 2008, \apss, 316, 113

 \bibitem[\protect\citeauthoryear{C\'orsico \&
     Althaus}{2006}]{corsico06} C\'orsico A. H., Althaus L. G., 2006,
   \aap, 454, 863

 \bibitem[\protect\citeauthoryear{Dziembowski}{1971}]{dziembowski71}
   Dziembowski W. A., 1971, {\em Acta Astron.}, 21, 289

 \bibitem[\protect\citeauthoryear{Fontaine et al.}{2008}]{fontaine08}
   Fontaine G., Brassard P., Green E. M., Chayer P., Charpinet S.,
   Andersen M., Portouw J., 2008, \aap, 486, L39

 \bibitem[\protect\citeauthoryear{Gautschy \&
     Althaus}{2002}]{gautschy02} Gautschy A., Althaus L. G., 2002,
   \aap, 382, 141

 \bibitem[\protect\citeauthoryear{Hirsch \& Heber}{2008}]{hirsch08}
   Hirsch H., Heber U., 2008, to appear in ASPC, vol. 392, Proceedings
   of the Hot Subdwarf Stars and Related Objects, p. 175

 \bibitem[\protect\citeauthoryear{Kawaler \&
     Bradley}{1994}]{kawaler94} Kawaler S. D., Bradley P. A., 1994,
   \apj, 427, 415

 \bibitem[\protect\citeauthoryear{Lawlor \&
     MacDonald}{2006}]{lawlor06} Lawlor T. M., MacDonald J. , 2006,
   \mnras, 371, 263

 \bibitem[\protect\citeauthoryear{Miglio et al.}{2008}]{miglio08}
   Miglio A., Montalb\'an J., Noels A., Eggenberger P., 2008, \mnras,
   386, 1487

 \bibitem[\protect\citeauthoryear{Moya, Garrido \& Dupret}{Moya et
     al.}{2004}]{moya04} Moya A., Garrido R., Dupret M.-A., 2004,
   \aap, 414, 1081

 \bibitem[\protect\citeauthoryear{Moya \& Garrido}{2008}]{moya08} Moya
   A., Garrido R., 2008, \apss, 316, 129

 \bibitem[\protect\citeauthoryear{Rodr\'\i guez-L\'opez et
     al.}{2009a}]{crl09sdss} Rodr\'\i guez-L\'opez C. et al., 2009,
   \mnras, accepted (astro-ph/0909.0930)

 \bibitem[\protect\citeauthoryear{Rodr\'\i guez-L\'opez et
     al.}{2009b}]{crl09} Rodr\'\i guez-L\'opez C., Moya A., Garrido
   R., MacDonald J., Oreiro R., Ulla A., 2009b, \mnras, accepted
   (astro-ph/0909.3778), Paper~I


 \bibitem[\protect\citeauthoryear{Smeyers et al.}{1995}]{smeyers95}
   Smeyers P., De Boeck I., Van Hoolst T., Decock L., 1995, \aap, 301,
   105

 \bibitem[\protect\citeauthoryear{Smeyers \& Moya}{2007}]{smeyers07}
   Smeyers P., \& Moya A., 2007, \aap,465, 509

 \bibitem[\protect\citeauthoryear{Str\"oer et al.}{2007}]{stroer07}
   Str\"oer A., Heber U., Lisker T., Napiwotzki R., Dreizler S.,
   Christlieb N., Reimers D., 2007, \aap, 462, 269

 \bibitem[\protect\citeauthoryear{Tassoul}{1980}]{tassoul80} Tassoul
   M., 1980, \apjs, 43, 469

 \bibitem[\protect\citeauthoryear{Winget, Van Horn \& Hansen}{Winget
     et al.}{1981}]{winget81} Winget D. E., Van Horn H. M., Hansen
   C. J. 1981, \apj, 245, L33

 \bibitem[\protect\citeauthoryear{Woudt et al.}{2006}]{woudt06} Woudt
   P. A. et al., 2006, \mnras, 371, 1497

\end{thebibliography}
\end{document}